\documentclass[twocolumn,showpacs,nofootinbib,amsmath,amssymb,prl,floatfix,superscriptaddress]{revtex4}

\usepackage{hyperref}
\usepackage{amsfonts}
\usepackage{graphicx}

\begin{document}


\title{Relativistic three-body effects in black hole coalescence}

\author{Manuela Campanelli} \affiliation{Department of Physics and Astronomy,
and Center for Gravitational Wave Astronomy, 
The University of Texas at Brownsville,
80 Fort Brown, Brownsville, TX 78520, USA}

\author{Miranda Dettwyler} \affiliation{Department of Physics and Astronomy,
and Center for Gravitational Wave Astronomy, 
The University of Texas at Brownsville,
80 Fort Brown, Brownsville, TX 78520, USA}

\author{Mark Hannam} \affiliation{Department of Physics and Astronomy,
and Center for Gravitational Wave Astronomy, 
The University of Texas at Brownsville,
80 Fort Brown, Brownsville, TX 78520, USA}

\affiliation{Theoretisch-Physikalisches Institut,
Friedrich-Schiller-Universit\"{a}t, 07743 Jena, Germany}

\author{Carlos O. Lousto} \affiliation{Department of Physics and Astronomy,
and Center for Gravitational Wave Astronomy, 
The University of Texas at Brownsville,
80 Fort Brown, Brownsville, TX 78520, USA}

\begin{abstract}
Three-body interactions are expected to be common in globular clusters
and in galactic cores hosting supermassive black holes. Here we consider
an equal-mass binary-black-hole system in
the presence of a third black hole. Using numerically
generated binary-black-hole initial-data sets, and first- and
second-order post-Newtonian (1PN and 2PN) techniques, we find that the
presence of the third black hole has non-negligible relativistic
effects on the location of the innermost stable circular orbit (ISCO),
and that these effects arise at 2PN order. In particular, we study 
the more astrophysically realistic situation of a stellar-mass black-hole
binary in orbit about a third supermassive black hole. In general, the
proximity of the massive black hole has stabilizing effects on the
orbiting binary, leading to an increase in merger time and 
a decrease of the terminal orbital
frequency, and an amplification of the gravitational radiation emitted
from the binary system by up to $6\%$ percent.
\end{abstract}

\pacs{04.25.Dm, 04.25.Nx, 04.30.Db, 04.70.Bw}

\maketitle

\section{Introduction}

Gravitational radiation, as predicted by general relativity, plays an
important observable role in the relativistic dynamics of
astrophysical systems. Collisions of compact objects, such as neutron
stars and black holes, produce characteristic gravitational wave
signals that are observable up to very high redshifts. In particular,
binary-black-hole systems are the prime scientific target source of
gravitational waves for both the current generation of ground-based
detectors, such as LIGO~\cite{LIGO}, and for the next generation of
space-based detectors, such as LISA~\cite{LISA1}.

Theoretically accurate models of black hole coalescences based
on the theory of general relativity are expected to provide crucial
information for the interpretation of these gravitational-wave
observations. So far, all numerical models have focused on isolated
binary-black-hole systems. However, in a realistic scenario it is
reasonable to expect that some of these black-hole binaries will, at
some point in their lifetime, interact gravitationally with a third
compact object. The presence of the third body may influence the
evolution and gravitational radiation emission of the
black-hole binary during the inspiral and merger phase. This may
produce some important observational effects on the gravitational
waveforms.

Relativistic three-body interactions are expected to play an important
role in astrophysical scenarios including (1) multibody interactions
in high-density cores of globular clusters
\citep{Gultekin:2003xd,miller02}, (2) stellar-mass binary-black-hole
systems interacting with a central supermassive black hole, and (3)
hierarchical triples of massive black holes that might be formed in
the nuclei of galaxies undergoing sequential mergers
\citep{makino90, valtonen96}. 
In the globular cluster scenario, a possible mechanism to produce
hierarchical triple systems is through binary-binary interactions
\citep{miller02,mikkola84,mcmillan91,rasio95}. At least 20-50\% of
binary-binary encounters may actually result in a stable hierarchical 
three-body system \citep{miller02}. Estimates of the lifetime of a triple 
system are made in
\citep{mcmillan91}, where it is noted that a three-body system can
survive several hundred times longer than the orbital period of one of
the original binaries \citep{makino90}. Further astrophysically motivated
studies of three-body systems have been considered in
Refs.~\citep{2004RMxAC..21..147V,1995MNRAS.273..751V,1995CeMDA..62..377P}.

In this paper we consider the effect of the presence of a third black 
hole on the location of the innermost stable circular orbit (ISCO) of a 
binary non-spinning black-hole system. We use numerical and post-Newtonian
techniques to study two three-body scenarios: (1) a test case, in which 
the third black hole has comparable mass to each black hole in the binary, 
and for simplicity is stationary, and (2) the more astrophysically 
realistic case of a stellar-mass black-hole binary in orbit about a 
supermassive black hole. Scenario (2) could be treated using only
post-Newtonian methods, but in scenario (1) we find that calculations at
2PN accuracy give results consistent with the fully general-relativistic
numerical approach.

\section{Three black-hole configuration}
\label{sec:configuration}

\begin{figure}[!ht]
\begin{center}
\includegraphics[scale=0.38]{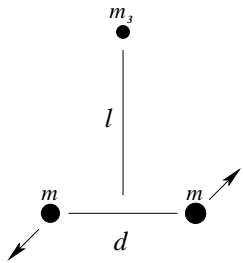}
\caption{Initial three black hole configuration: the third body sits 
motionless directly above the orbital plane of the binary.}
\label{fig:configuration}
\end{center}
\end{figure}

We consider the following configuration, illustrated in
Fig.~\ref{fig:configuration}. Two black holes are separated by a
distance $d$, and are assumed to be in quasi-circular orbit. A third
black hole is located on the axis of rotation of the binary, a
distance $l$ from the center of rotation. Each black hole in the
binary has equal and opposite momentum $P$, and the total angular
momentum of the binary is $J = Pd$. In the first set of results we will 
treat the third black hole as being instantaneously stationary; we will later
relax this simplifying assumption.

Each black hole in the binary has mass $m$. For the remainder of
this paper, when referring to numerical calculations, $m$ will
correspond to the bare mass of each black hole, as defined later; in 
analytic post-Newtonian calculations $m$ will be the Newtonian mass of the
body. The third body has mass $m_3$. The total mass of the 
three black holes will be $M = 2m + m_3$. In numerical calculations, the
total ADM mass of the spacetime (which will differ from $M$), will be
denoted by $M_{ADM}$.

We will model this configuration using two approaches. The first
is to consider numerically generated initial-data sets,
i.e., solutions of the initial-value equations of general
relativity. We will identify circular orbits (and the ISCO) of the binary 
by searching for minima in an effective-potential of the spacetime
\citep{cook94}. As such, we are making a number of assumptions and
approximations.  We are assuming that the effective-potential method
reliably yields parameters consistent with quasi-circular orbits, and
that the initial data we use are a reasonable approximation to the
``correct'' initial data for two black holes in orbit. In this numerical
approach we are ignoring
entirely any motion of the third black hole: we consider only the
effect of a stationary third black hole on the instantaneous binding
energy of the binary system, and make the loose assumption that its
effect on the locations of quasi-circular orbits will be indicative of
the effect of a third black hole passing near the binary.
 The purpose of this work is to determine
the qualitative effect of the presence of a third body on the ISCO of
a binary. 

The second approach is to use post-Newtonian
calculations. Quasi-circular orbits and the ISCO have been identified
for binary systems (see, for example, \citep{damour00,blanchet02}), and it 
is possible to extend this 
procedure to a three-body system. The results, obtained to 1PN and 2PN 
accuracy, are compared with the numerical initial-data results. The
post-Newtonian approach has the advantage that it is
straightforward to generalize to cases where the binary is in circular
orbit about the third black hole.

\section{Method}
\subsection{Initial-data approach}

We first outline the numerical initial-data approach. 
Black-hole initial data consist of the solutions to the four initial-value
equations that result from the (3+1) decomposition of the field equations 
of general relativity \citep{adm,york79}. These equations are the 
Hamiltonian and momentum constraints for the spatial metric $\gamma_{ij}$ 
and extrinsic curvature $K_{ij}$ on one time-slice \citep{york79}, which, 
in vacuum, are
\begin{eqnarray}
\bar{\nabla}_j \left(K^{ij} - \gamma^{ij}K \right) & = & 0, \label{eqn:mc1} \\
R+K^2-K_{ij}K^{ij} & = & 0, \label{eqn:hc1}
\end{eqnarray} where $K = \gamma^{ij} K_{ij}$ is the trace of the 
extrinsic curvature, $R$ is the Ricci scalar, and all quantities, 
including the covariant derivative $\bar{\nabla}_i$, are defined with 
respect to the time-slice. 

One way to solve the constraint equations is through the conformal
transverse-traceless decomposition \citep{york79}. In this decomposition
there exist analytic solutions of the momentum constraint, the Bowen-York 
solutions, that can describe any number of boosted black holes. In particular,
we can write down a solution for two boosted black holes, plus a single 
stationary black hole (which is a trivial solution). 
The Hamiltonian constraint remains to be solved numerically, and we do 
this using the puncture approach of Brandt and Br\"{u}gmann \cite{brandt97}. 

Given the freedom in specifying the orbital parameters of the
initial-data sets for black-hole spacetimes, we would like to identify
which of those sets represents two black holes in quasi-circular
orbits.  To this end we implement the effective-potential method of
Cook \citep{cook94} in which minima are located in sequences of total
ADM energy, $E_{ADM}$, versus coordinate separation, $d$, for constant
values of the orbital angular momentum of the binary, $J$, and individual
black-hole masses $m$ and $m_3$.  In this work the ADM mass calculated at each 
puncture is used to denote the bare mass of each black hole 
\cite{baker02b}. This procedure is carried out for many sequences until the
lowest value of $J$ that produces a sequence with a minimum is
located.  We denote this last minimum as the innermost stable circular
orbit (ISCO) of the binary.  Once the ISCO is located, its orbital
angular frequency, $\Omega$, is calculated via
\begin{equation}
\Omega = \frac{\partial E_{ADM}}{\partial J}\Bigg|_{m_i}.
\label{eqn:omega}
\end{equation} The application of this approach for equal-mass 
binary-black-hole systems can be found in \citep{cook94,baum00}. 

\subsection{Post-Newtonian approach}

We now turn to the procedure to find quasi-circular orbits and the ISCO 
in post-Newtonian theory. The ISCO for two-body systems has been calculated 
using post-Newtonian methods by a number of authors
\citep{damour00,blanchet02}. The approach can be generalized to three
bodies and compared with our numerical results. This method will not tell us
anything about the stability of the ``orbits'' that have been identified, and indeed
in general we do not expect a general relativistic three-body system to be any
more stable than its Newtonian counterpart. However, we can still identify 
parameters that meet the requirements of quasi-circular orbits, as described
below, and suggest configurations that we reasonably expect to be stable. In particular,
when a stellar-mass black-hole binary orbits a third supermassive black hole, which 
is the astrophysically relevant scenario we ultimately wish to study. 

In the two-body case, the post-Newtonian and numerical (Bowen-York data) 
results differ, but in this work we are interested in the differential effect 
of the presence of a third body, and can test whether that effect is
comparable when calculated using numerical or post-Newtonian methods.

We locate the post-Newtonian ISCO using the procedure outlined in
\citep{damour00}. We start with the Hamiltonian for a three-body
system in the configuration described above, which is given by
\citep{ohta74,schafer85,schafer87}. The three-body Hamiltonian is a function
of the locations, masses and momenta of the three bodies; in total, 21 
parameters. In the configuration described above, these parameters are
reduced to five: $m$, the mass of each body in the equal-mass binary, $m_3$, the 
mass of the third body, $d$, the binary separation, $l$, the distance of the 
third body from the center of mass of the binary, and $p$, the mangnitude
of the linear momentum of each body in the binary. In all cases we 
consider, the third body is either stationary, or its motion is such that
its angular momentum about the origin of the system is perpendicular to the
angular momentum of the binary, and so the magnitude of the binary's 
total angular momentum may be clearly given as $J = pd$.

To identify circular orbits, one first locates
minima order by order in the post-Newtonian expansion of the Hamiltonian. 
This procedure was performed semi-analytically with {\it Mathematica}, to 
yield the separation $d$ of a circular orbit to 2PN accuracy in terms of the 
total binary angular momentum, $J$. This value of $d$ was inserted into the 
post-Newtonian Hamiltonian to get the energy, $E_{circ}$, of circular orbits to
1PN or 2PN accuracy and to calculate the angular velocity of the orbit via
(\ref{eqn:omega}), with $E_{ADM}$ now replaced by $E_{circ}$. 
This procedure was performed for different values of $J$, to produce a
plot of $E_{circ}$ versus $\Omega$; the minimum in this plot
corresponds to the ISCO.

\section{Results}

\begin{figure}[!ht]
\begin{center}
\includegraphics[scale=0.6]{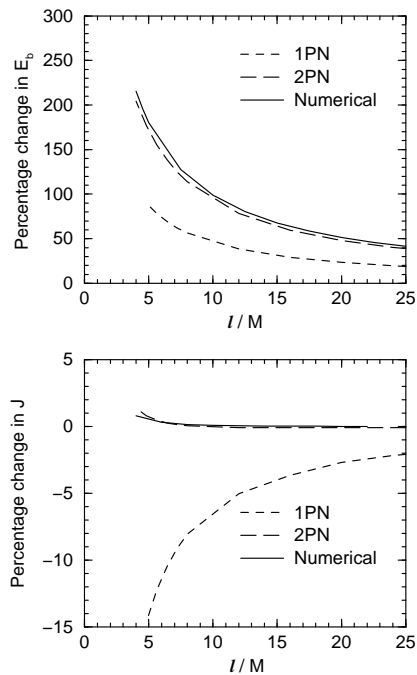}
\caption{The percentage change in the binding energy of the three-body system,
$E_b$, and the total angular momentum of the binary, $J$, at the binary's ISCO,
as a function of the coordinate distance of a third body from the
center of the binary. $m = 1$; $m_3 = 0.5$, and the third body
is stationary.}
\label{fig:Compare_05}
\end{center}
\end{figure}

In the first configuration, the masses are $m = 1$; $m_3 = 0.5$, 
and the third body is stationary. The results of the numerical and 
post-Newtonian approaches are compared in Fig.~\ref{fig:Compare_05}, where 
shows the percentage change in the binding energy of the system, $E_b$, 
and the total angular momentum of the binary, $J$, at the binary's ISCO, as a
function of the coordinate distance of a third body from the center of
the binary is shown. The percentage differences are with respect to the
corresponding two-body ISCO for numerical, 1PN and 2PN
calculations. Note that the numerical, 1PN and 2PN results are in
different gauges, and the coordinate distance of the third black hole,
$l/M$, may correspond to (slightly) different physical distances in each of the
three gauges. It is therefore not appropriate to make a quantitative
comparison of these results. Note also that we are considering the binding
energy of the entire three-body system, not only the binary. However, we can 
make the qualitative observations that (1) the binding energy of the system 
increases dramatically as the third black hole is placed closer to the binary,
(2) there is better agreement between 2PN and numerical data than
between 1PN and numerical data, and (3) most significantly, we need at least
2PN accuracy to see the qualitatively correct effect on the binary's
angular momentum, $J$. For this reason we will use 2PN data in the subsequent
calculations for a better motivated astrophysical scenario. 

Having seen that the presence of a third body has a significant effect
on the binary's ISCO, we wish to consider the effect in an
astrophysically more realistic situation: the third body is far larger
than each of the bodies in the binary (by a factor of, for example,
$10^5$), and the binary is in orbit about the third body. It is not
practical to consider this scenario numerically: in particular, we
cannot achieve suitable accuracy for such extreme mass ratios with
our finite-difference code. However, the previous results show that 2PN
calculations are adequate to describe the problem.

\begin{figure}[!ht]
\begin{center}
\includegraphics[scale=0.6]{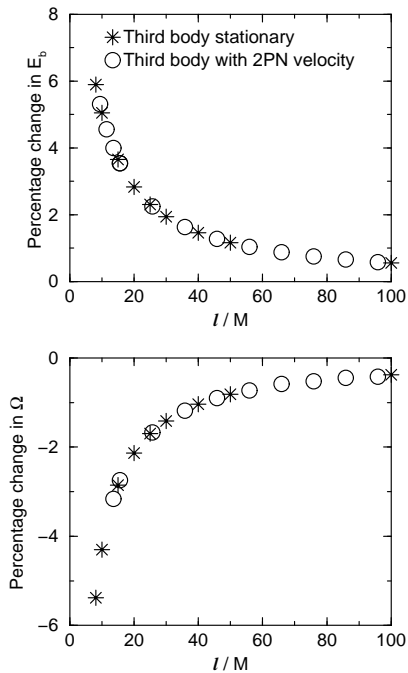}
\caption{The percentage change in the binding energy, $E_b$, and 
orbital frequency, $\Omega$, of the binary at the
ISCO, as a function of the distance of a third body from the center of
the binary. The masses are $m_1 = m_2 = 1$; $m_3 = 10^5$.}
\label{fig:Compare_SM}
\end{center}
\end{figure}

Fig.~\ref{fig:Compare_SM} shows the percentage change in the binding
energy of the binary due to the presence of a third body with a mass
$10^5$ times larger than that of each body in the binary. This
configuration models a stellar-mass binary orbiting a supermassive
black hole in a galactic core. All results were calculated to 2PN
accuracy. The effect of the location of the third body was calculated
when the third body was stationary (starred points), and when it had
momentum (circles) consistent with the 2PN circular orbit of a
two-body system of masses $2m$ and $m_3$ respectively. We can
see that the effect is practically the same in both cases. Note that
we are now considering the binding energy of the binary only (using the
two-body Hamiltonian with the binary parameters found in the orbit search
procedure), not that of the entire system, as we did in 
Fig.~\ref{fig:Compare_05}, and the effect of the third body is much smaller. 
However, it is still appreciable, around 6\%, when the third body is close 
to its own ISCO with the binary.

To examine the generality of this effect, we placed the third body 
at different orientations to the binary: in the 
plane of the binary, at 45 degrees to that plane, and we 
changed the direction of the third body's momentum. The results were 
unaffected by these changes to within the accuracy of the method.

In addition to being an interesting purely relativistic effect,
the displacement of the ISCO due to the presence of the third body
has observational consequences on the emission of gravitational
waves. A larger negative value of the binding energy at the ISCO can
be associated with a larger loss of energy radiated by the system
during the inspiral phase (comparable to that during the plunge and
ring-down phase \cite{baker01,baker02}.) We thus expect (1) an
increase in the terminal amplitude of the inspiral gravitational
waveform; (2) an increase in the duration of the pre-plunge phase; and
(3) since the orbital frequency of the ISCO decreases due to the
presence of the third body (see Fig.~\ref{fig:Compare_SM}), the
corresponding waveform will display a lower pre-plunge frequency.
Similarly, the rotation parameter of the post-plunge remnant black
hole will be smaller, due to a decrease in $J_{ISCO}$. This, in turn,
will reduce the value of the least damped quasi-normal frequency of the
final Kerr black hole.

\acknowledgments 
We wish to thank Marc Freitag, David Merritt and Cole Miller for helpful
discussions, and for carefully reading this manuscript.  The authors
gratefully acknowledge the support of the NASA Center for
Gravitational Wave Astronomy (CGWA) at The University of Texas at Brownsville
(NAG5-13396), and NSF grants PHY-0140326 and PHY-0354867. Numerical
results were obtained on the CGWA {\it Funes} cluster.

\bibliographystyle{apsrev}
\bibliography{3bh}
\thebibliography{PRL}

\end{document}